# A Model of Random Multiple Access in Unlicensed Spectrum Systems

Dmitriy K. Kim
High School of Digital Technology
Narxoz University
Zhandosov Str. 55, Almaty, Kazakhstan
Email: dmitriy.kim@narxoz.kz

Georgi D. Georgiev
Center for Energy Solutions
3K Solar Corporation
Razlog Str. 705, Varna, Bulgaria

E-mail: georgidg@gmail.com

Natalya V. Markovskaya
Institute of Information Systems
and Information Security
Saint-Petersburg State University
of Aerospace Instrumentation
B. Morskaya 67, Saint Petersburg, Russia
Email: mnv-k52@guap.ru

*Abstract*— We consider classical multiple access system with a single transmission channel, finite number of users (users) and randomised transmission protocol (ALOHA). We assume that every user sends messages to the base station with various intensities. Due to the overlapping of messages during their sending there are restrictions on the time between the messages and a mathematical model adequate to the physical one becomes quite complicated. In this work we propose a simplified mathematical model that is easy to analyze and takes into account the properties of real systems.

Keywords — mathematical model; random multiple access, ALOHA algorithm; LoRaWAN: Poisson stream.

## I. Introduction

Increasing growth of information content in our world requires faster processing and transmission which leads to the development of data transmission systems. One of the main goals is to efficiently arbitrate access of users to shared network. Furthermore, over last few decades we may observe active growth of data transmission systems based on multiple access channels such as radio channels and satellite communication channels. Among these access control methods that allow a number of users access methods to a shared network, a special mention is deserved by random multiple access methods with conflict resolution. Widespread application of random multiple access is due to its use in the Ethernet and wireless networks over IEEE 802.11 (Wi-Fi). Today, random multiple access with conflict resolution is also used for backup of shared channel in regional wireless networks over IEEE 802.16. One of the main properties of random multiple access is lower delay for moderate arrival rates when compared to other methods of multiple access.

Moreover, wireless LPWAN networks are also becoming more and more widespread [1]. They are used to transmit information about the results of measurements at objects located over a large area. The LoRaWAN technologies as part of the LPWAN make it possible to create low-cost data collection and transmission systems. There is no need to build and maintain a complex multi-domain network. Network devices share a common medium for transmitting data. To avoid the synchronization effect when transmitting messages, some LoRaWAN equipment manufacturers provide the facility to set a random interval between messages (see, for example, [1]). Therefore, many LPWANs, including the LoRa technology, use the ALOHA algorithm to resolve conflicts in message transmission [1]-[6].

The classical multiple access system with a single transmission channel and the randomised transmission protocol (ALOHA) is well known (see, for example, [7]). In the classical random multiple access model with ALOHA algorithm for conflict resolution there are $N$ users, each of whom sends messages to the base station at random points in time. If during the transmission of a message from one user the other user begins to transmit, then there is a collision and the messages are lost and do not reach the base station.

The distribution of time between sent messages is the same for all users and independent of each other. We may consider the special case when the messages sent from each user form a Poisson point process with intensity $\lambda$, i.e. time between sent messages is exponential with parameter $\lambda$. This simplification facilitates the construction of the mathematical model and obtaining conclusions from it. In particular, in this case it is easy to find the distribution of the total message stream from all users, which will be again Poisson's with intensity

$$\Lambda = N * \lambda.$$

Moreover, if each message is successfully delivered to the base station with probability $p$, then for the Poisson case the stream of successfully delivered messages from each node will again be Poisson with the parameter $\Lambda p$. In this work, we consider a generalization of the classical model, assuming that each $i$-th user sends messages to the base station with a various intensity $\lambda_i$. This approach allows us to take into account the specificity of the transmitted data when monitoring the parameters of some system over time. If some user transmits a message with not very important information, it can send messages with low intensity, in contrast to the user, information from which about the state of the system is very important and therefore the transmission from it occurs with high intensity.

Moreover, in real information transmission systems, there are technical limitations on the time between sending messages from each user in unlicensed spectrum systems, for example in LoRa and other LPWAN networks. In general terms, this time should be much longer than the transmission time (airtime) of a single message [8]. This also means that the user cannot send a new message until it has finished sending the previous one.

The goal of our work is to suggest a mathematical model for message transmission using the ALOHA algorithm by users with different intensities. The model should take into account the properties of real systems as the constraint on the time between sending messages and allow to obtain analytical results as distribution for the stream of successfully delivered messages from each node at least for Poisson input stream. The first part of the work a mathematical model adequate to the physical model of message transmission is considered. In the second part, we propose a simplified mathematical model for Poisson point process that reflects the main features and limitations of real model and allows studying characteristics of the system.

## II. MODEL DESCRIPTION

Let us consider with $N$ users. Denote the random intervals between messages from user $i = 1, 2, ..., N,$ by non-negative random variables $\tau_n^{(i)}, \ n = 1, 2, ...,$ with $\mathbf{E}\left|\tau_n^{(i)}\right| < \infty$. Put that $\tau_n^{(i)}$ are independent sequences of independent equally distributed random variables. If $\tau_n^{(i)}, \ i = 1, 2, ..., N,$ have an exponential distribution, we obtain a Poisson stream of messages. Define their intensities as

$$\lambda_i = 1/\mathbf{E}\tau_n^{(i)}.$$

Put $Q > 0$ as airtime. According to [8] the air time must be less than 1% of the time between sent messages, i.e.

$$\tau_1^{(i)} \geq 100Q. \qquad (1)$$

Consider a point process of total message stream from all users. Denote by $\tau_n$ the intervals between messages. They are independent identically distributed random variables. If we remove from total message stream all messages, distances between which do not exceed $2Q,$ then we obtain a total stream of successfully delivered messages (see Fig. 1) with intervals between messages $\delta_n$.

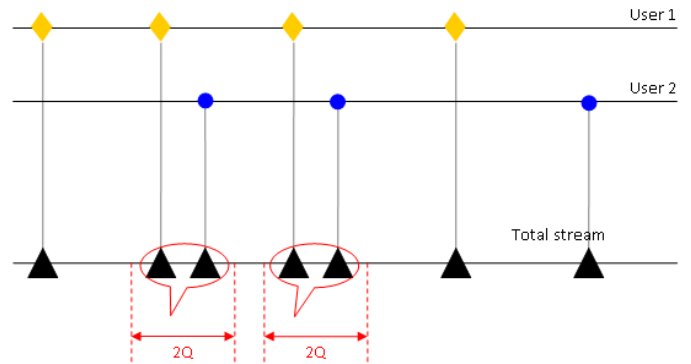

Fig. 1. Input streams of two users (yellow diamonds and blue circles) and their total stream of messages (black triangles) in time. Messages with distance less $2Q$ are removing.

Due to Markov property $\delta_n$ has distribution as

$$\begin{aligned}&\tau_1 I(\tau_1 > Q, \tau_2 > Q) + \\ &(\tau_1 + \tau_2)I(\tau_1 < Q, \tau_2 > Q, \tau_3 > Q) + \\ &(\tau_1 + \tau_2 + \tau_3)I(\tau_1 < Q, \tau_2 < Q, \tau_3 > Q, \tau_4 > Q) + ...\end{aligned} \qquad (2)$$

Now, we can return each message to the user who sent it from the total stream of successfully delivered messages. So we find a stream of successfully delivered messages for each user (see Fig. 2).

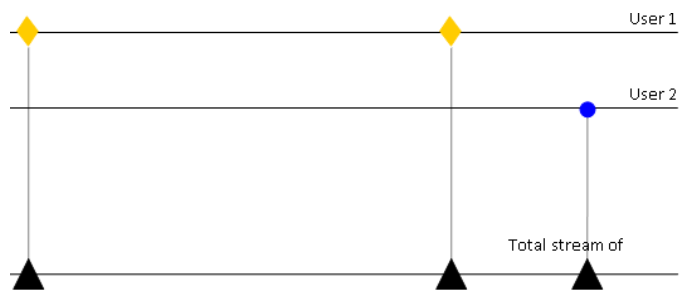

Fig. 2. Stream of successfully delivered messages of two users (yellow diamonds and blue circle) and their total stream of successfully delivered messages (black triangles) in time.

It is easy to see that this model is correct but not convenient to apply analytical methods and obtain explicit conclusions. For example it is difficult to find explicit distribution for the stream of successfully delivered messages from each node.

## III. SIMPLIFIED MATHEMATICAL MODEL

In this section we will simplify the mathematical model from the previous paragraph by removing or replacing constraints (1) and (2). Suppose that $\tau_n^{(i)}$, $i = 1, 2, ..., N$, have exponential distribution with parameters $\lambda_i > 0$, i.e. for any $x > 0$

$$\mathbf{P}(\tau_l^{(i)} > x) = e^{-\lambda_i x}.$$

Instead of (1) we put that the message transmission time (airtime $Q$) does not exceed 1% of the average time between sent messages, i.e.

$$\mathbf{E}\tau_l^{(i)} \geq 100Q.$$

or

$$\lambda_i \leq \frac{1}{100Q}.$$

This means that we neglect the restriction of (1) because it is inconvenient to work with shifted distributions. Recall that in real systems a user cannot start a transmission until the previous one is finished. We replace the strict restriction (1) on the time between messages by the same restriction, only for the average time between messages.

Then the total message stream will also be Poisson (superposition theorem) with intensity

$$\Lambda = \lambda_1 + \lambda_2 + ... + \lambda_N$$

and the intervals between messages $\tau_n$ in the total stream will be independent exponentially distributed random variables with parameter $\Lambda$.

Then, as before, we remove all messages from the total message stream, the distances between which do not exceed $2Q$. That is, instead of the restriction (1) we consider only restriction (2) on the intervals between messages in the total message stream. In the next step we change this restriction as well. Let $\tilde{\tau}_n$ be independent copies of $\tau_n$. Now, instead of random variables $\delta_n$ (see (2)) we introduce new random variables

$$\tilde{\delta}_n = \tau_1 I(\tilde{\tau}_1 > 2Q) + (\tau_1 + \tau_2)I(\tilde{\tau}_1 < 2Q, \tilde{\tau}_2 > 2Q) + (\tau_1 + \tau_2 + \tau_3)I(\tilde{\tau}_1 < 2Q, \tilde{\tau}_2 < 2Q, \tilde{\tau}_3 > 2Q) + ....$$

It is easy to see that the random variables $\delta_n$ have an exponential distribution with the parameter

$$\Lambda e^{-2Q\Lambda}.$$

Now let us formulate our simplified mathematical model. Consider the general message stream as a Poisson process with parameter $\Lambda$. Suppose that each message from the total stream is successfully delivered with probability $p = e^{-2Q\Lambda}$ to the base station and leaves the system with probability $1-p$. In this case, the interval between successfully delivered messages $\tilde{\delta}_n$ will have an exponential distribution with the parameter $\Lambda p$ and the total flow of successful messages will be Poisson.

So, the requirement on the length of the interval between sent messages we replace by the condition that each message is delivered to the base station with probability $p = e^{-2Q\Lambda}$. Consider now the average time between successfully delivered messages. It is easy to see that

$$\frac{1}{\Lambda e^{-2Q\Lambda}} \geq \frac{1}{\Lambda}(1 + 2Q\Lambda) = 2Q + \frac{1}{\Lambda}.$$

It means that in our simplified model the average time between successfully delivered messages for the total stream of messages is bigger than $2Q$. Recall that, in Section II, to obtain the total stream of successfully delivered messages, we removed all messages whose interval length was less than $2Q$. In our model, we allow any interval length between sent messages in the total message stream, but the average time between successfully delivered messages is greater than $2Q$.

Now assign each message to the $i$-th user with probability

$$\frac{\lambda_i}{\Lambda}, i = 1, 2, ..., N.$$

As a result, if every message from total stream of messages belong to $i$-th user with probability $\frac{\lambda_i}{\Lambda}, i = 1, 2, ..., N.$ we obtain a Poisson stream of successfully delivered messages of each user with intensity [9]

$$\Lambda e^{-2Q\Lambda} \frac{\lambda_i}{\Lambda} = \lambda_i e^{-2Q\Lambda}.$$

Remark that for equal

$$\lambda_i = \lambda$$

we obtain that

$$\lambda_i e^{-2Q\Lambda} = \lambda e^{-2NQ\lambda}.$$

As before, it is easy to see that the average time between successfully delivered messages from the $i$-th node is also greater than $2Q$.

$$\frac{1}{\lambda_i e^{-2Q\Lambda}} \geq 2Q\frac{\Lambda}{\lambda_i} + \frac{1}{\lambda_i}.$$

IV. CONCLUSION

The multiple access system with a single transmission channel, finite number of users (users) and randomised transmission protocol (ALOHA) is considered. Each of $N$ users sends messages to the base station at random times. If during the transmission of a message from one user any user begins to transmit, then there is a collision and the messages are lost and do not delivered the base station. In the classical model, each user independently of the others sends messages according to a Poisson distribution of the same intensity. We are studying the model that, unlike the classical case, allows each user to send messages with various intensity. This approach allows us to take into account the specifics of the transmitted data, for example, when monitoring a system with many users. If the monitoring of a particular user has priority over the others, it can send messages with a higher intensity than the others. In addition, in real systems there are constraints on the time between sent messages: the duration of message sending (airtime) must not be more than 1% of the time between sent messages. Moreover, in reality, each user cannot start a new message transmission until the current message transmission is finished. The study of such a problem requires the construction of an adequate and at the same time simple mathematical model, which would allow to obtain in an explicit form the distribution of the stream of successfully delivered messages of each user and all of them together.

According to the classical model with the total Poisson input message stream, each message is successfully delivered with probability $p$, or is lost with probability $1-p$. So the stream of successfully delivered messages is also Poisson with the parameter $\Lambda p$. In our model, we have shown why and under what assumptions we can choose $p = e^{-2Q\Lambda}$.

In real systems, a user cannot start the next transmission until the previous one is incomplete, and that if during one user's message transmission, of length $Q$, another user starts transmitting a message, all messages are lost. We have replaced these requirements in our model by assuming that the length of the interval between messages has an exponential distribution with a mean greater than $2Q$.

Note that an exponential distribution between messages and $p = e^{-2Q\Lambda}$ are used in works on transmission systems in unlicensed spectrum systems (see [10]) without mathematical foundation. In our work we show that this approach has the right to exist and give a mathematical model to prove it.

Firstly, we define a general Poisson stream of successfully delivered messages in which there is no restriction on the interval between messages, but there is a dependence of its average length on airtime. Then, based on the property of the Poisson process, we determine the stream of successfully delivered messages of each user.